\documentclass[8pt,tighten,preprint2,flushrt]{aastex}
\usepackage{graphicx}
\usepackage{geometry}
\usepackage{float}
\usepackage{epsfig,lscape} \usepackage[dvipdfm]{hyperref} \usepackage{natbib}
\usepackage{slashbox}
\usepackage{multirow}
\usepackage{subfigure}
\usepackage{color}
\usepackage{color,soul}
\usepackage{rotating}
\newcommand{\myemail}{zouhu@nao.cas.cn}

\begin{document}
\footnotesize

\title{South Galactic Cap $u$-band Sky Survey (SCUSS): Data Release}
\author{Hu Zou \altaffilmark{1}, Xu Zhou \altaffilmark{1}, Zhaoji Jiang \altaffilmark{1}, Xiyan Peng \altaffilmark{1}, Dongwei Fan \altaffilmark{1}, Xiaohui Fan \altaffilmark{2}, Zhou Fan \altaffilmark{1}, Boliang He\altaffilmark{1}, Yipeng Jing \altaffilmark{3}, Michael Lesser \altaffilmark{2}, Cheng Li \altaffilmark{4}, Jun Ma \altaffilmark{1}, Jundan Nie \altaffilmark{1}, Shiyin Shen \altaffilmark{4}, Jiali Wang \altaffilmark{1}, Zhenyu Wu \altaffilmark{1}, Tianmeng Zhang \altaffilmark{1}, Zhimin Zhou \altaffilmark{1}}
\altaffiltext{1}{Key Laboratory of Optical Astronomy, National Astronomical Observatories, Chinese Academy of Sciences, Beijing, 100012, China; \myemail}
\altaffiltext{2}{Steward Observatory, University of Arizona, Tucson, AZ 85721, USA}
\altaffiltext{3}{Center for Astronomy and Astrophysics, Department of Physics and Astronomy, Shanghai Jiao Tong University, Shanghai 200240, China}
\altaffiltext{4}{Shanghai Astronomical Observatory, Chinese Academy of Sciences, Shanghai 200030, China}

\begin{abstract} 
	The South Galactic Cap $u$-band Sky Survey (SCUSS) is a deep $u$-band imaging survey in the south Galactic cap using the 2.3m Bok 
telescope. The survey observations were completed at the end of 2013, covering an area of 
about 5000 square degrees. We release the data in the region with an area of about 4000 deg$^2$ 
that is mostly covered by the Sloan digital sky survey. The data products contain
calibrated single-epoch images, stacked images, photometric catalogs, and a catalog of 
star proper motions derived by \citet{pen15}. The median seeing and magnitude limit ($5\sigma$) 
are about 2\arcsec.0 and 23.2 mag, respectively. There are about 8 million objects having  
measurements of absolute proper motions. All the data and related documentations can be 
accessed through the SCUSS data release website \url{http://batc.bao.ac.cn/Uband/data.html}.
\end{abstract}

\keywords{surveys --- catalogs --- techniques: image processing --- techniques: photometric}

\section{Introduction}
The South Galactic Cap $u$-band Sky Survey (SCUSS) is an international 
cooperative project between the National Astronomical Observatories of 
Chinese Academy of Sciences (NAOC) and the Steward Observatory of the University 
of Arizona (X. Zhou et al. 2015, in preparation). The survey was originally planned 
to perform a sky survey of about 3700 deg$^2$ in the south Galactic 
cap by using the 2.3m Bok telescope. The project 
was initiated in fall 2009 and its first run started in 2010 September. 
The survey ended its observation in 2013 December. The final survey area 
is about 5000 deg$^2$, far beyond the planned area.

The main goal of the survey is to supply a $u$-band catalog for the spectroscopic 
target selection of the Large Sky Area Multi-Object Fiber Spectroscopy Telescope \citep{cui12}. 
Besides, combined with the $g, r, i$, and $z$-band data of the Sloan Digital Sky Survey
\citep[SDSS;][]{yor00}, the deep SCUSS $u$-band data can be used to study 
the Milk Way and galaxies. A series of papers based on the SCUSS data have been 
published, including investigating the halo structure of the Galaxy \citep{nie15}, 
calculating star proper motions \citep{pen15}, estimating the Galactic photometric 
metallicity and model parameters \citep{jia14,guj15}, and selecting spectroscopic 
targets, such as quasars and emission line galaxies \citep{com15,rai15,zou15b}. 

This paper describes the data set of the SCUSS data release that is made publicly 
available. The paper is organized as follows: Section \ref{sec2} summarizes the survey and 
data reduction; Section \ref{sec3} presents the data products including the calibrated images 
and photometric catalogs; Section \ref{sec4} gives an analysis of the data quality. Section \ref{sec5} 
describes the catalog of star proper motions derived by \citet{pen15}; Section \ref{sec6} 
is the conclusion.

\section{The Survey and Data Reduction} \label{sec2}
The SCUSS is a wide and deep $u$-band sky survey in the south Galactic cap. The survey 
uses the 90 inch (2.3m) Bok telescope that belongs to the Steward Observatory. 
It operates every night of the year except Christmas Eve and 
the maintenance period in August. The camera, named 90Prime, is installed at the prime focus 
(correct focal ratio $f$/2.98). It contains four 4k$\times$4k backside-illuminated CCDs that 
are assembled in a 2$\times$2 array with gaps along both vertical and horizontal directions. 
The CCDs are optimized for the $u$-band response, giving a quantum efficiency close to 80\%. 
The edge-to-edge FOV is about 1\arcdeg.08$\times$1\arcdeg.03. The adopted filter 
is similar to the SDSS $u$ band. The SCUSS $u$ filter is somewhat bluer and narrower. The  
central wavelength and FWHM are 3538 and 520 \AA, respectively \citep{zou15a}.

\begin{figure*}
	\includegraphics[width=\textwidth]{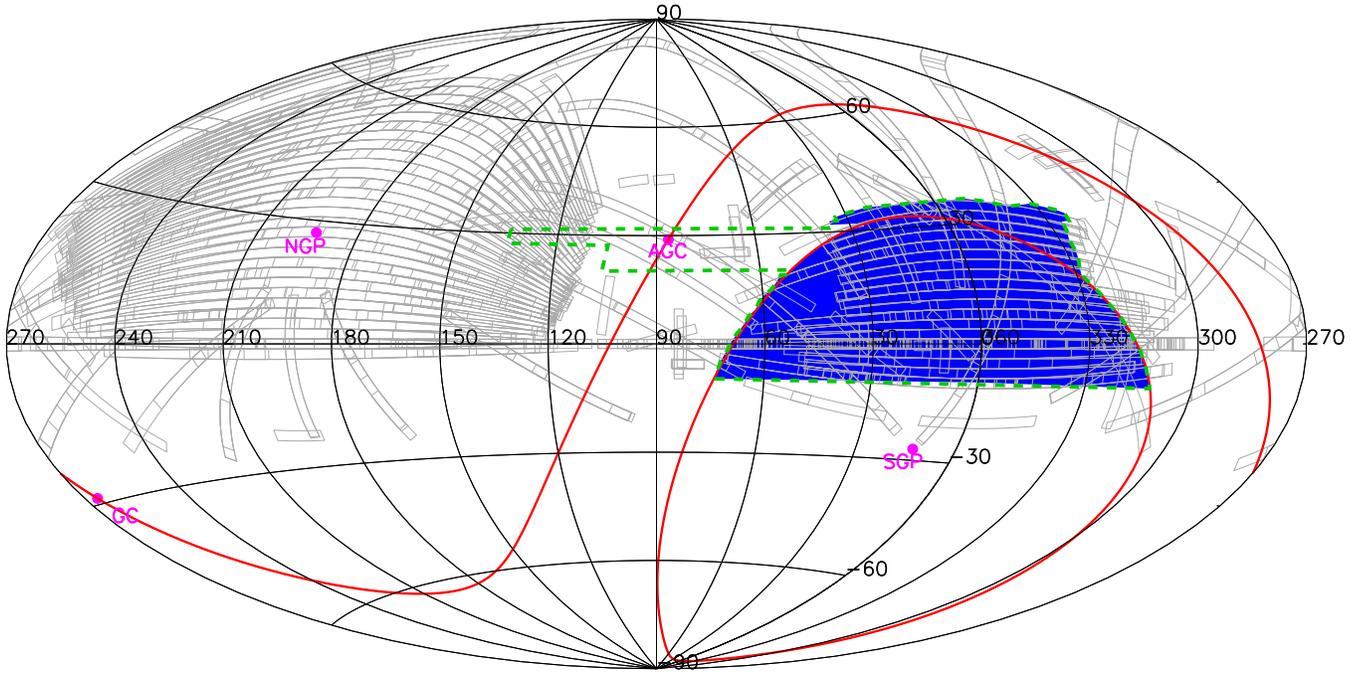}
	\caption{Aitoff projection of the SCUSS footprint. This projection is 
		centered at ($\alpha=90\arcdeg$, $\delta=0\arcdeg$). Pink filled 
		circles show the south Galactic pole (SGP), north Galactic pole (NGP), 
		Galactic center (GC), and anti-Galactic center (AGC). The red solid 
		curves show the Galactic plane and Galactic latitude of -30\arcdeg. 
		The green dashed lines present the actual coverage of the survey, whose 
		area is about 5000 deg$^2$. The blue filled region is the area where 
		the data to be released in this paper. The SDSS imaging runs are also 
		overlapped in gray. 
	\label{fig1}}
\end{figure*}

The originally designed survey footprint is located within the region of $\delta> -10\arcdeg$ and Galactic latitude 
$b < -30\arcdeg$. The observation started in 2010 September and ended in 2013 December. 
The final area is about 5000 deg$^2$ (dashed green line in Figure \ref{fig1}), including 
the planned area, an extra area in the northwest corner, and the region extending to the 
anti-Galactic center. In this paper, we only release the data shown in the blue area of Figure \ref{fig1}. The area is about 4000 deg$^2$, 92\% of 
which is covered by the SDSS. Each field has two continuous exposures, giving a total 
exposure time of 5 minutes. These 
two exposures are dithered by 1/2 of the CCD size, which benefits the internal photometric 
calibration and gap filling. In this way, most of the field is covered by two exposures. Some gap areas 
are covered by one exposure. The exposure time of 5 minutes generates an expected depth of 
23.0 mag. 

The following is a summary of the data reduction. More details can be found in \citet{zou15a}.

(1) Detrending: a dedicate image processing pipeline was compiled, which performs some 
standard calibrations (overscan subtraction, bias correction, and flat-fielding, etc.) 
and special handling (crosstalk, CCD artefacts, abnormal overscan, and sky gradient etc.) 

(2) Astrometry: the UCAC4 catalogs are used to derive the astrometric solutions. The global external position error is about 0\arcsec.13. The average internal astrometric error, from sources on overlapping exposures is about 0\arcsec.09.

(3) Magnitude calibrations: the SDSS catalogs are used to make external photometric 
calibrations. We calculate the photometric zeropoints (ZPs) separately for the four amplifiers to 
reduce the effect of gain and photometric response variations. For images out of the SDSS coverage, 
we interactively derive their ZPs internally by using the common stars 
in the overlapped area. 

(4) Image stacking: single-epoch images with specified qualities are stacked. More than 91\%
of stacked images are assembled by using single-epoch images with consistent qualities. 

(5) Photometry: SExtractor photometry \citep{ber96} with automatic elliptical apertures is performed on 
stacked images. Aperture, point spread function (PSF) and model magnitudes are measured 
from both stacked images and co-added flux measurements on single-epoch images. The model 
photometry uses the SDSS $r$-band shape parameters. Due to improper flat-fielding, 
scattered light, and focal plane distortion, the photometric ZP would vary with the 
position in the CCD plane so that there are photometric residuals across the CCD. We derive 
such residual maps for all CCDs and use them to correct the magnitudes. 

\section{Data Products} \label{sec3}
\subsection{Calibrated images}
Single-epoch images are calibrated by the dedicated image processing pipeline. There are a 
total of 44,937 images. The coordinate system adopts the ARC celestial projection, mostly 
used in Schmidt plate astrometry, with a 2nd-order radial distortion. The coordinate 
transformation between the focal plane and the celestial coordinates can be made by using 
our IDL/Python programs\footnote{\url{http://batc.bao.ac.cn/BASS/lib/exe/fetch.php?media=scuss:single-epoch_image_calibration:a8_convert.tar.gz}} with the 8 coefficients in the FITS header (keywords of 
A81, A82, ..., and A88) and a 2nd-order coefficient, which is implemented as inline functions in the programs. The usages 
of these programs can be found in the webpage\footnote{\url{http://batc.bao.ac.cn/BASS/doku.php?id=scuss:single-epoch_image_calibration:idl_python_programs}}. The WCS parameters in the header are incorrect. 
The photometric ZPs of four amplifiers are presented in keywords of CALIA731, 
CALIA732, CALIA733, and CALIA734, which are corresponding to the northeast, southeast, 
northwest, and southwest quadrants of the image. Thus, the magnitude can be calculated as 
$m = -2.5\mathrm{log}_{10} F + 25 + \mathrm{ZP}$, where $F$ is the measured flux in DN. 
To ensure homogeneity of the imaging depth and coverage completeness of the stacked images, 
we flag single-epoch images according to their qualities. The images with 
seeing $<$ 3\arcsec.0, sky ADU $<$ 500, and ZP $>$ 3.5 are flagged with ``1." If a sky region 
has not enough images flagged with ``1." the corresponding images in this region with seeing 
$<$ 3\arcsec.0 are flagged with ``2." If the region still has not enough images to meet the 
depth requirement, the rest of the images located in this region are flagged with ``3.". 
For the remaining images, we flag the ones with seeing $<$ 3\arcsec.0 with ``4" or otherwise with ``5." 
Table \ref{tab1} presents main keywords in the FITS header of the single-epoch images.

\begin{table*}
\centering
\caption{Main Keywords in the FITS header of Single-epoch Images and their Meanings\label{tab1}}
\begin{tabular}{lll}
\hline
\hline
Keyword & Data Type & Meaning \\
\hline
RA-OBS  & string &  R.A. of the field center in J2000 \\
DEC-OBS & string & Decl. of the field center in J2000 \\
CCD\_NO & int &  CCD number   \\                                           
DATE-OBS & string & UTC date when the shutter was opened   \\                          
TIME-OBS & string & UTC time when the exposure was started   \\                    
EXPTIME  & float & Exposure time (seconds)                \\        
RA  & string &  Right ascension in the specified epoch    \\                  
DEC  &  string & Decl. in the specified epoch         \\          
HA     & string & Hour angle                                     \\
EPOCH  & float & Equinox for R.A. and Decl.              \\            
FILTER  & string & Filter name                  \\
OBJECT & string & Field name                                     \\                     
RDNOCAL & string & Calculated readout noises for four CCDs \\                                  
GAINCAL & string & Calculated gains for four CCDs   \\
SKYADU  & float    &  Sky background in DN      \\
SEEING  & float &   Seeing in arcsec     \\
A81     & double &  Coefficients for coordinate transformation        \\
A82     & double &  Coefficients for coordinate transformation       \\
A83     & double &  Coefficients for coordinate transformation        \\
A84     & double &  Coefficients for coordinate transformation        \\
A85     & double &  Coefficients for coordinate transformation          \\ 
A86     & double &  Coefficients for coordinate transformation         \\
A87     & double &  Coefficients for coordinate transformation         \\              
A88     & double &  Coefficients for coordinate transformation         \\                               
AIRMASS & float &   Airmass when the exposure was taken             \\
MAZIMUTH & float & Moon azimuth in degrees from south through west       \\
MALITIUD & float & Moon altitude  in degrees                    \\      
MANGLE  & float & Position angle of Moon relative to the camera center     \\  
CALIA73  &    float & Zeropoint for the whole CCD image   \\
CALIA731 &   float & Zeropoint for the Amp. \#1. of the CCD image   \\
CALIA732 &   float & Zeropoint for the Amp. \#2. of the CCD image   \\   
CALIA733 &   float & Zeropoint for the Amp. \#3. of the CCD image   \\   
CALIA734 &   float & Zeropoint for the Amp. \#4. of the CCD image   \\  
\hline
\end{tabular}
\end{table*}

There are a total of 3700 stacked images, each of which has an area of 
1\arcdeg.08$\times$1\arcdeg.04 and about 1{\arcmin} overlaps with adjacent images. 
These stacked images are assembled by the single-epoch images flagged as ``1," ``2," and ``3." 
The coordinate system is a purely linear transformation in the ARC celestial projection. The ZP 
is stored in the keyword of CALIA73. The other header keywords are similar to those in 
the single-epoch images, but the WCS parameters are accurate enough. In addition, there 
is a weight map corresponding to each stacked image, giving the number of exposures. 

\subsection{Photometric Catalogs}
The catalogs contain both magnitudes measured on stacked images and co-added 
magnitudes from measurements on single-epoch images. The objects come from 
the SCUSS detections and SDSS catalogs with any of the $ugriz$ magnitudes 
in DR9 brighter than 23.5 mag. The matching error between the SCUSS and SDSS 
is 2\arcsec.0. In the catalogs, SDSS objects can be recognized by the NUMBER column, 
where NUMBER $<$ 49,000 or 50,000 $<$ NUMBER $<$ 60,000. The SCUSS unique objects 
have NUMBER $>$ 60,000. Extra matched fainter SDSS objects within 2{\arcsec} have 
49,000 $<$ NUMBER $<$ 50,000.

The SExtractor photometry is performed only on stacked images, providing the 
automatic magnitude, Kron radius, shape parameters, object classification, etc. 
Aperture, PSF, model magnitudes are measured on both stacked images and 
single-epoch images. Co-added magnitudes are 
derived from these measurements on single-epoch images. Flags for the PSF magnitude (column PSFFLAG) are coded in 
decimal and expressed as a sum of powers of 2: 1 for CCD artefacts; 2 for bad 
pixels; 4 for saturated pixels; 8 for contamination from neighbors; 16 for 
edges of the image. The co-added flag (PSFADDFLAG) is the combination of corresponding 
flags of the same object measured on multiple single-epoch images.

If stack images are combined with single images of similar qualities, 
the magnitudes measured on stacked images are better than the co-added ones, 
since the object number is 20\% higher. Conversely, the co-added magnitudes should 
be better. We can refer to \citet{zou15a}  for the general guidelines to use the magnitudes. 
Table \ref{tab2} and \ref{tab3} show both the SCUSS and SDSS photometric  
information included in the catalogs. All magnitudes are corrected to the aperture 
magnitudes (7\arcsec.26 in radius), which is also used for photometric calibrations. 
\begin{table*}
\centering
\caption{Main columns in SCUSS photometric catalogs\label{tab2}}
\begin{tabular}{lll}
\hline
\hline
Field & Data type & Meaning \\
\hline
   NUMBER         &LONG    &ID of objects in stacked images \\
   RA2000         &STRING  &R.A. in J2000\\
   DEC2000        &STRING  &Decl. in J2000\\
   X              &FLOAT   &Image X of SDSS objects\\
   Y              &FLOAT   &Image Y of SDSS objects\\
   BER\_X          &FLOAT   &SExtractor X of objects detected on stacked images\\
   BER\_Y          &FLOAT   &SExtractor Y of objects detected on stacked images\\
   MAG\_AUTO       &FLOAT   &Automatic magnitude derived by SExtractor\\
   MAGERR\_AUTO    &FLOAT   &Automatic magnitude error derived by SExtractor\\
   KRON\_RADIUS    &FLOAT   &Kron radius in pixels derived by SExtractor\\
   MAG\_PETRO      &FLOAT   &Petrosian magnitude derived by SExtractor\\
   MAGERR\_PETRO   &FLOAT   &Petrosian magnitude error derived by SExtractor\\
   PETRO\_RADIUS   &FLOAT   &Petrosian radius in pixels derived by SExtractor\\
   FLUX\_RADIUS    &FLOAT   &Half-light radius in pixels derived by SExtractor\\
   FWHM\_IMAGE     &FLOAT   &FWHM of objects in pixels derived by SExtractor\\
   BERTIN\_G\_S    &FLOAT  &Stellarity (0 galaxy; 1 star) derived by SExtractor\\
   A\_AXIS         &FLOAT   &Length of the major axis in pixels derived by SExtractor\\
   ELLIPTICITY    &FLOAT   &Ellipticity derived by SExtractor\\
   THETA          &FLOAT   &Position angle in degrees derived by SExtractor\\
   BERTIN\_CLASS   &INT     &SExtractor Flags \\
   COMBINE\_NUMB    &INT     &Exposure number in the stacked image \\
   PSFMAG         &FLOAT   &PSF magnitudes on the stacked image\\
   PSFERR         &FLOAT   &PSF magnitude error on the stacked image\\
   PSFFLAG        &INT     &Flags of the PSF magnitude on the stacked image\\
   APMAG          &DOUBLE  &Aperture magnitude on the stacked image (12 apertures)\\
   APMAGERR       &DOUBLE  &Aperture magnitude error on the stacked image\\
   MODELMAG       &FLOAT   &Model magnitude on the stacked image\\
   MODELMAGERR    &FLOAT   &Model magnitude error on the stacked image \\
   PSFADD         &DOUBLE  &Co-added PSF magnitude from single-epoch images\\
   PSFADDERR      &DOUBLE  &Co-added PSF magnitude error from single-epoch image\\
   PSFADDSTD      &DOUBLE  &Standard deviation of the co-added PSF magnitude\\
   PSFADDFLAG     &LONG    &Flags of the co-added PSF magnitude \\
   PSFADDNUM      &INT     &Exposure number for the co-added PSF magnitude\\
   APADD          &DOUBLE  &Co-added aperture magnitudes (12 apertures)\\
   APADDERR       &DOUBLE  &Co-added aperture magnitude errors\\
   APADDSTD       &DOUBLE  &Standard deviations of the co-added aperture magnitudes\\
   APADDNUM       &LONG    &Exposure numbers for the co-added aperture magnitudes\\
   MODELADD       &FLOAT   &Co-added model magnitude\\
   MODELADDERR    &FLOAT   &Co-added model magnitude error.\\
   MODELADDSTD    &FLOAT   &Standard deviation of the co-added model magnitude \\
   MODELADDNUM    &INT     &Exposure number for the co-added model magnitude\\
   JDMEAN         &DOUBLE  &Average Julian day for each object \\
   MATCH\_ERR      &FLOAT   &Match error in arcsec between SCUSS detected objects and SDSS objects\\
\hline
\end{tabular}
\end{table*}

\begin{table}
\tiny
\centering
\caption{Main SDSS columns included in SCUSS photometric catalogs \label{tab3}}
\begin{tabular}{lll}
\hline
\hline
Field & Data type & Meaning \\
\hline
   SDSSOBJID      &STRING  &SDSS OBJID in SDSS DR9\\
   SDSSTYPE       &STRING  &SDSS object type (s: star; g: galaxy)\\
   PSFMAG\_U       &FLOAT   &SDSS $u$-band PSF magnitude\\
   PSFMAG\_G       &FLOAT   &SDSS $g$-band PSF magnitude\\
   PSFMAG\_R       &FLOAT   &SDSS $r$-band PSF magnitude\\
   PSFMAG\_I       &FLOAT   &SDSS $i$-band PSF magnitude\\
   PSFMAG\_Z       &FLOAT   &SDSS $z$-band PSF magnitude\\
   PSFMAGERR\_U    &FLOAT   &SDSS $u$-band PSF magnitude error\\
   PSFMAGERR\_G    &FLOAT   &SDSS $g$-band PSF magnitude error\\
   PSFMAGERR\_R    &FLOAT   &SDSS $r$-band PSF magnitude error\\
   PSFMAGERR\_I    &FLOAT   &SDSS $i$-band PSF magnitude error\\
   PSFMAGERR\_Z    &FLOAT   &SDSS $z$-band PSF magnitude error\\
   PETROMAG\_U     &FLOAT   &SDSS $u$-band Petrosian magnitude\\
   PETROMAG\_G     &FLOAT   &SDSS $g$-band Petrosian magnitude \\
   PETROMAG\_R     &FLOAT   &SDSS $r$-band Petrosian magnitude\\
   PETROMAG\_I     &FLOAT   &SDSS $i$-band Petrosian magnitude\\
   PETROMAG\_Z     &FLOAT   &SDSS $z$-band Petrosian magnitude\\
   PETROMAGERR\_U  &FLOAT   &SDSS $u$-band Petrosian magnitude error\\
   PETROMAGERR\_G  &FLOAT   &SDSS $g$-band Petrosian magnitude error\\
   PETROMAGERR\_R  &FLOAT   &SDSS $r$-band Petrosian magnitude error\\
   PETROMAGERR\_I  &FLOAT   &SDSS $i$-band Petrosian magnitude error\\
   PETROMAGERR\_Z  &FLOAT   &SDSS $z$-band Petrosian magnitude error\\
   MODELMAG\_U     &FLOAT   &SDSS $u$-band model magnitude\\
   MODELMAG\_G     &FLOAT   &SDSS $g$-band model magnitude\\
   MODELMAG\_R     &FLOAT   &SDSS $r$-band model magnitude\\
   MODELMAG\_I     &FLOAT   &SDSS $i$-band model magnitude\\
   MODELMAG\_Z     &FLOAT   &SDSS $z$-band model magnitude\\
   MODELMAGERR\_U  &FLOAT   &SDSS $u$-band model magnitude error\\
   MODELMAGERR\_G  &FLOAT   &SDSS $g$-band model magnitude error\\
   MODELMAGERR\_R  &FLOAT   &SDSS $r$-band model magnitude error\\
   MODELMAGERR\_I  &FLOAT   &SDSS $i$-band model magnitude error\\
   MODELMAGERR\_Z  &FLOAT   &SDSS $z$-band model magnitude error\\
   CMODELMAG\_U    &FLOAT   &SDSS $u$-band Cmodel magnitude \\
   CMODELMAG\_G    &FLOAT   &SDSS $g$-band Cmodel magnitude\\
   CMODELMAG\_R    &FLOAT   &SDSS $r$-band Cmodel magnitude\\
   CMODELMAG\_I    &FLOAT   &SDSS $i$-band Cmodel magnitude\\
   CMODELMAG\_Z    &FLOAT   &SDSS $z$-band Cmodel magnitude\\
   CMODELMAGERR\_U &FLOAT   &SDSS $u$-band Cmodel magnitude error\\
   CMODELMAGERR\_G &FLOAT   &SDSS $g$-band Cmodel magnitude error\\
   CMODELMAGERR\_R &FLOAT   &SDSS $r$-band Cmodel magnitude error\\
   CMODELMAGERR\_I &FLOAT   &SDSS $i$-band Cmodel magnitude error\\
   CMODELMAGERR\_Z &FLOAT   &SDSS $z$-band Cmodel magnitude error\\
   EXTINCTION\_U   &FLOAT   &SDSS $u$-band extinction\\
   EXTINCTION\_G   &FLOAT   &SDSS $g$-band extinction\\
   EXTINCTION\_R   &FLOAT   &SDSS $r$-band extinction\\
   EXTINCTION\_I   &FLOAT   &SDSS $i$-band extinction\\
   EXTINCTION\_Z   &FLOAT   &SDSS $z$-band extinction\\
\hline
\end{tabular}
\end{table}
\section{Data Quality and Depth} \label{sec4}
\begin{figure*}
	\includegraphics[width=\textwidth]{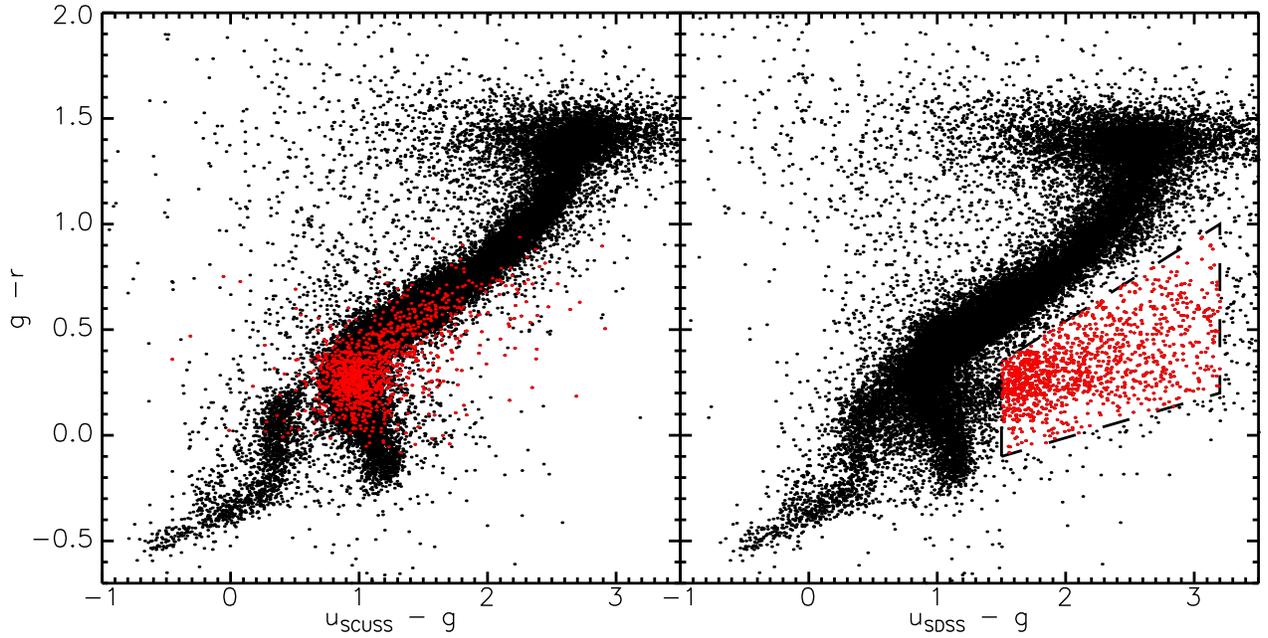}
	\caption{Left: stars spectroscopically identified by the SDSS in the color-color 
		diagram of $u_\mathrm{SCUSS} - g$ vs. $g - r$. Right: the same stars in the 
		color-color diagram of $u_\mathrm{SDSS} - g$ vs. $g - r$. The magnitudes are 
		corrected for the Galactic extinction \citep{sch98} and the color term used by 
		\citet{zou15b}. The dashed polygon indicates that the objects were initially 
		selected as quasar candidates but identified by the SDSS as stars. 
		Both the $g$ and $r$-band magnitudes come from the SDSS.  
	\label{fig2}}
\end{figure*}

\begin{figure*}
	\includegraphics[width=\textwidth]{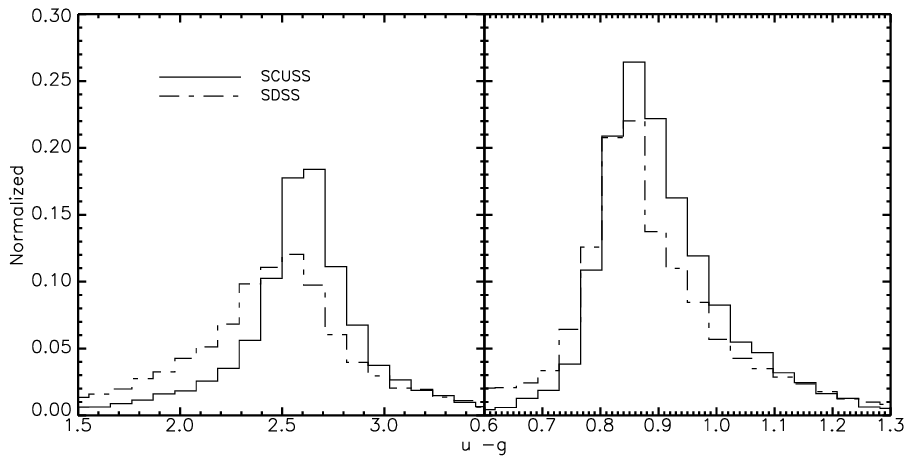}
	\caption{Left: the $u - g$ distribution of stars with $1.3 < g - r < 1.6$. Right: the $u - g$ 
		distributions of stars with $0.0 < g - r < 0.3$. The solid and dashed histograms 
		use the SCUSS and SDSS $u$ bands, respectively. 
	\label{fig3}}
\end{figure*}

The color-color diagram is an excellent tool to compare the photometry of the SCUSS 
and SDSS. Figure \ref{fig2} shows the scatters of stars in the plane of $u - g$ vs. $g - r$. 
These stars are spectroscopically identified by the SDSS. The SCUSS co-added PSF 
magnitude and SDSS PSF magnitudes are used for comparison. In this figure, the star 
sequence using the SCUSS $u$ band is tighter. We select two color intervals of $1.3 < g - r < 1.6$ 
(mostly M type stars) and $0.0 < g - r < 0.3$ (mostly A type stars) to show the $u - g$ distributions, which 
is  presented in Figure \ref{fig3}. The dispersion of the $u_\mathrm{SCUSS} - g$ color 
is smaller than that of the $u_\mathrm{SDSS} - g$. Moreover, objects located in the 
lower right (enclosed by a polygon in Figure \ref{fig2}) are identified as 
stars but initially selected as quasar candidates with redshift larger than 3.0. These stars were faint and  
mistakenly selected as quasars due to the bad SDSS $u$-band photometry. But they are still located 
in the star sequence when the SCUSS $u$ band is used. 

\begin{figure*}
	\includegraphics[width=\textwidth]{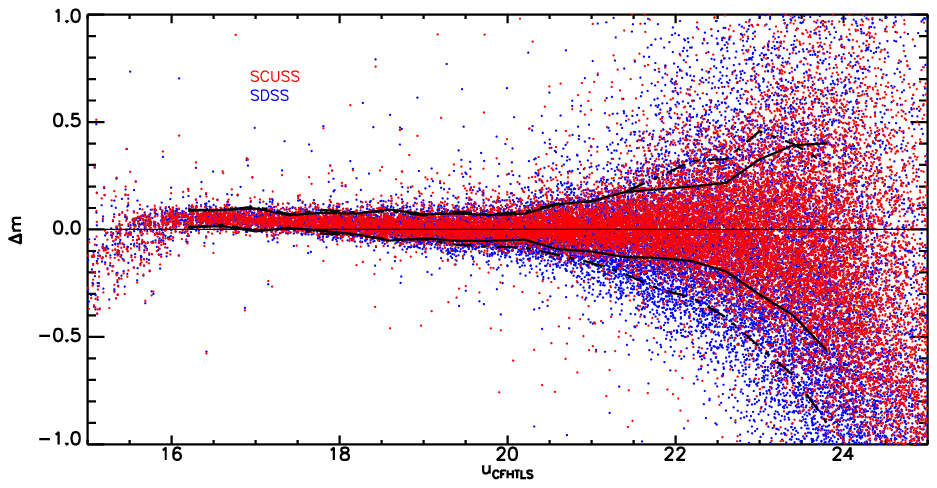}
	\caption{The $u$-band magnitude difference between the SCUSS/SDSS and the CFHTLS 
	as a function of the CFHTLS $u$-band magnitude. The blue points stand for the SDSS and the 
	red ones stand for the SCUSS. The solid and dashed lines show the RMS around the average 
	as a function of the CHFTLS $u$-band magnitude for the SCUSS and SDSS, respectively. 
	\label{fig4}}
\end{figure*}

The SCUSS and SDSS $u$-band photometry is also compared by using the catalogs from the 
Canada-France-Hawaii Telescope Legacy Survey (CFHTLS), whose wide-field $u$-band depth reaches  
about 25.2 mag (80\% completeness limit). We select the CFHTLS W4 field 
($\alpha = \mathrm{22^h13^m18^s}, \delta = \mathrm{+01^d19^m00^s}$) that is 
fully covered by the SCUSS. Figure \ref{fig4} shows the photometric comparison of 
point sources that are classified by the SDSS. The SCUSS and SDSS $u$-band PSF magnitudes 
are converted to the CHFTLS photometric system \footnote{http://cfht.hawaii.edu/Instruments/Imaging/MegaPrime/generalinformation.html}. The solid and dashed lines in this figure show 
the photometric RMS around the average offset as a function of the magnitude. 
The SDSS has a much larger offset when the magnitude is fainter. The magnitude offset 
between the SDSS and the CFHTLS at $u = 23.5$ is about 0.2 mag, while the one between 
the SCUSS and the CFHTLS is about -0.03 mag. In addition, for the same RMS of 0.2, the 
SCUSS and SDSS magnitude limits are about 22.6 and 21.4 mag, respectively. The SCUSS 
$u$ band is 1.2 mag deeper.

\begin{figure*}
	\subfigure{\includegraphics[width=0.5\textwidth]{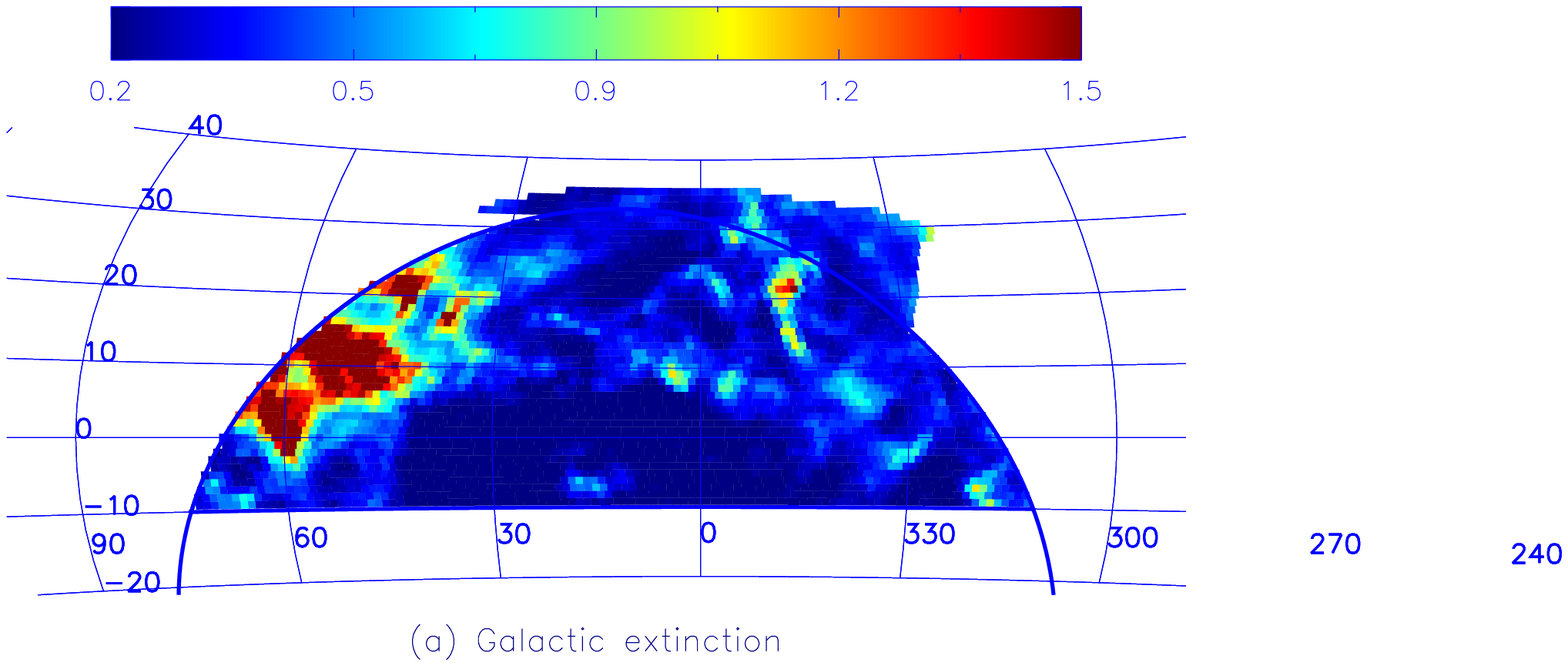}}
	\subfigure{\includegraphics[width=0.5\textwidth]{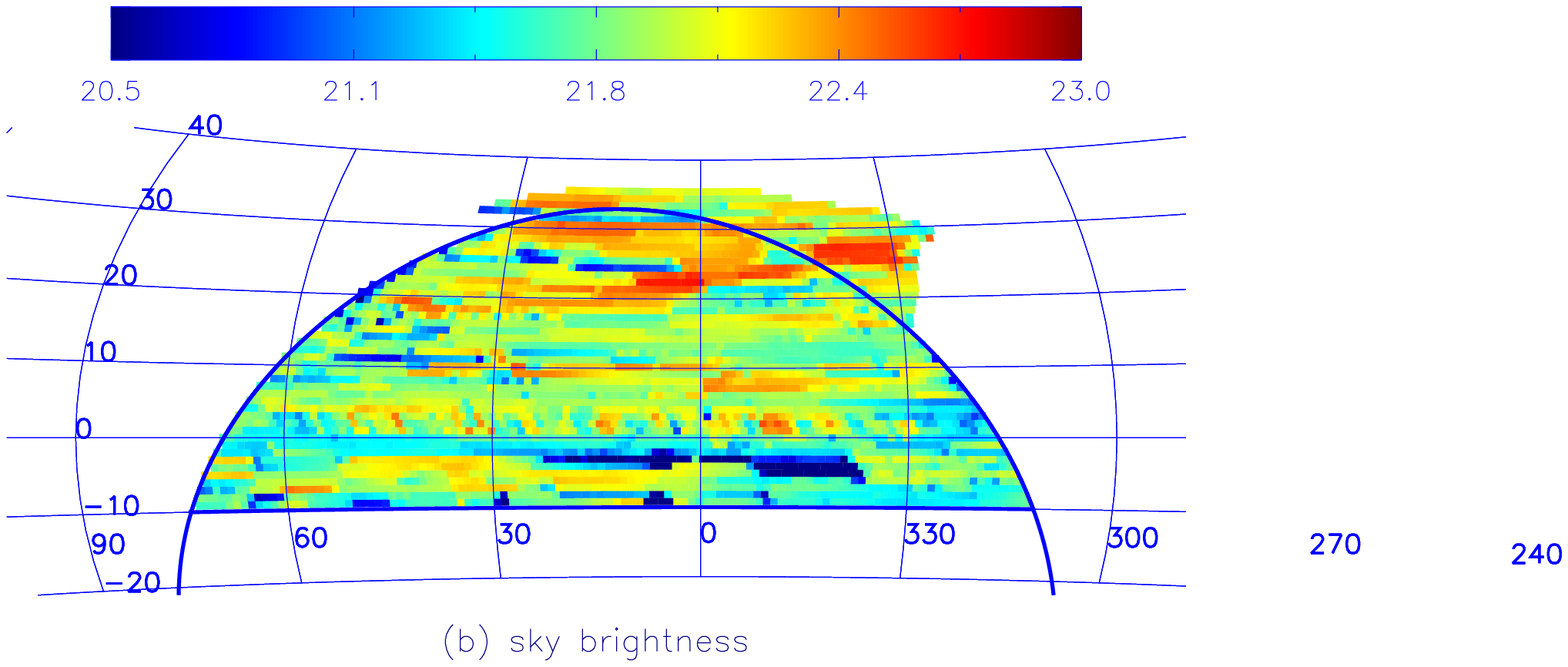}}
	\subfigure{\includegraphics[width=0.5\textwidth]{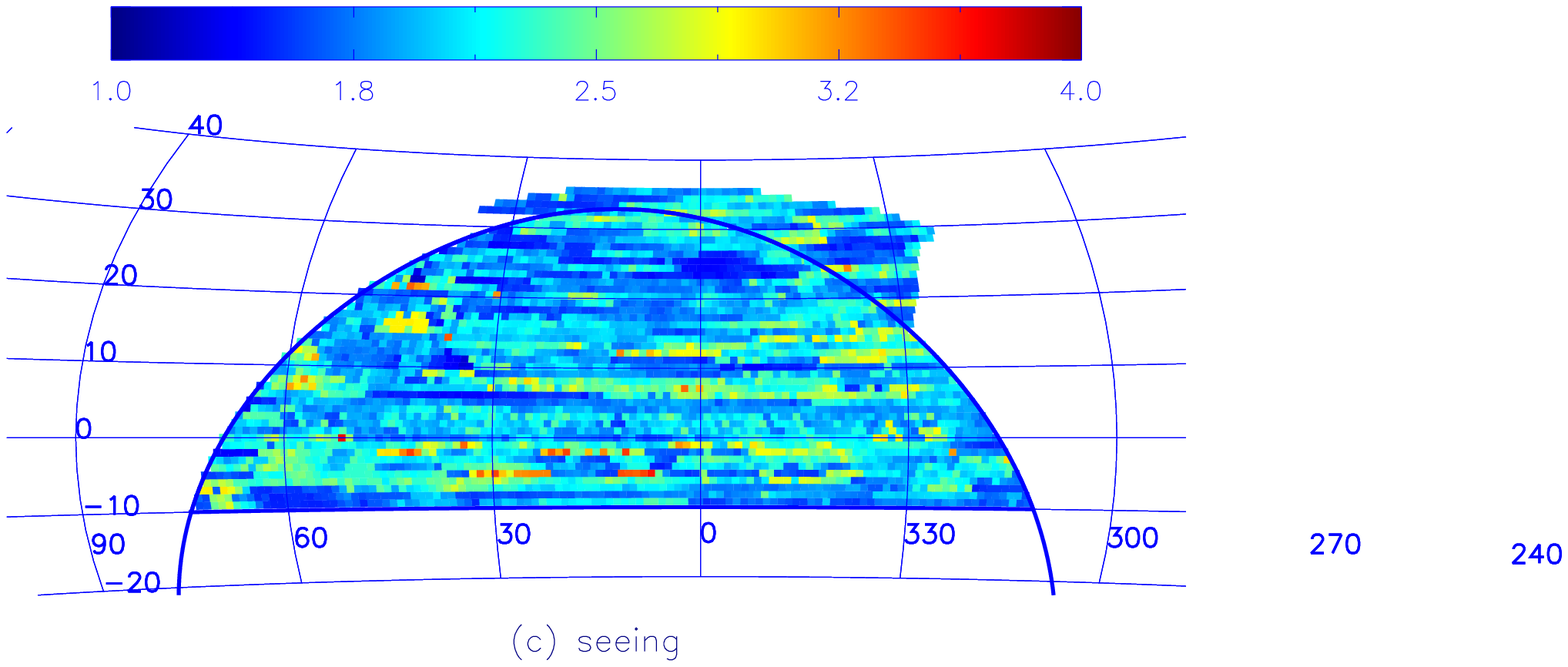}}
	\subfigure{\includegraphics[width=0.5\textwidth]{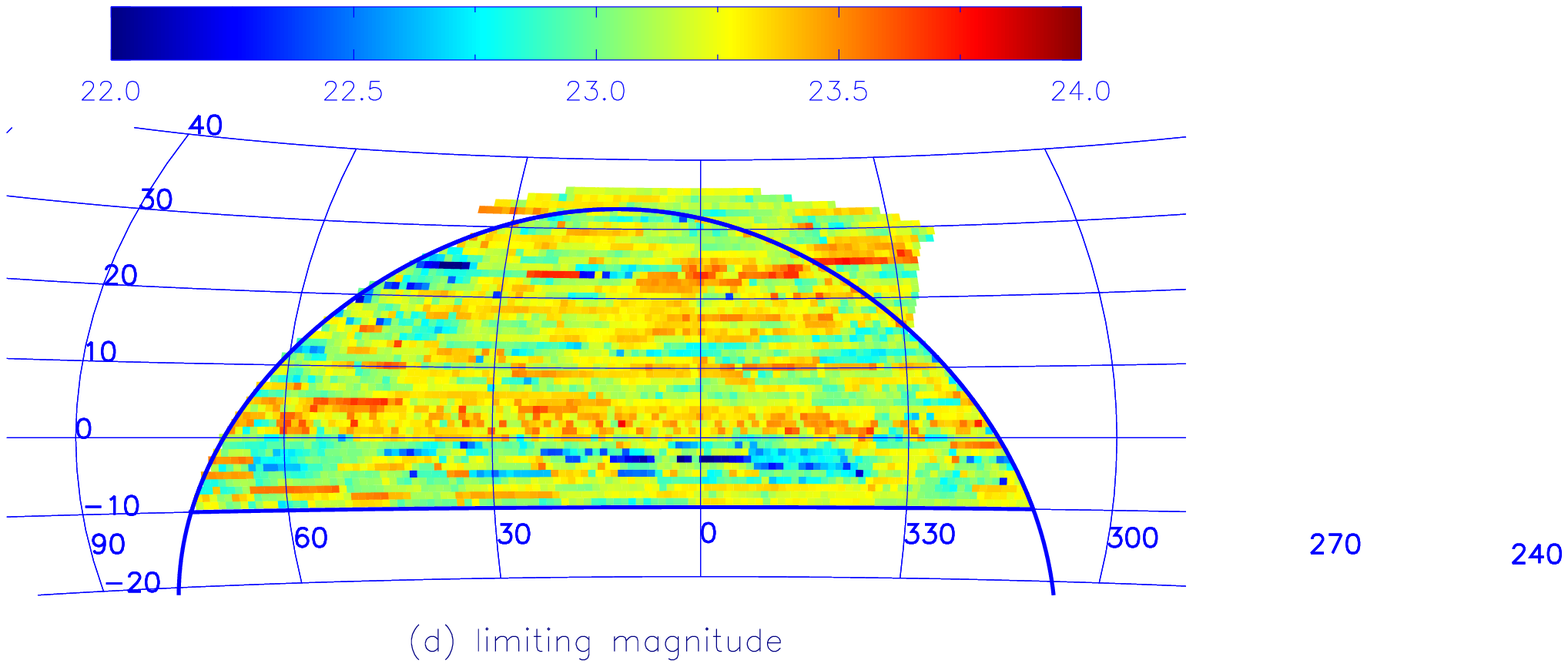}}
	\caption{(a): $u$-band Galactic extinction map in the SCUSS footprint; (b) sky brightness map in 
	mag arcsec$^{-2}$. (c) Seeing map in arcsec; (d) magnitude limit map in mag. \label{fig5}}
\end{figure*}

\begin{figure*}
	\includegraphics[width=\textwidth]{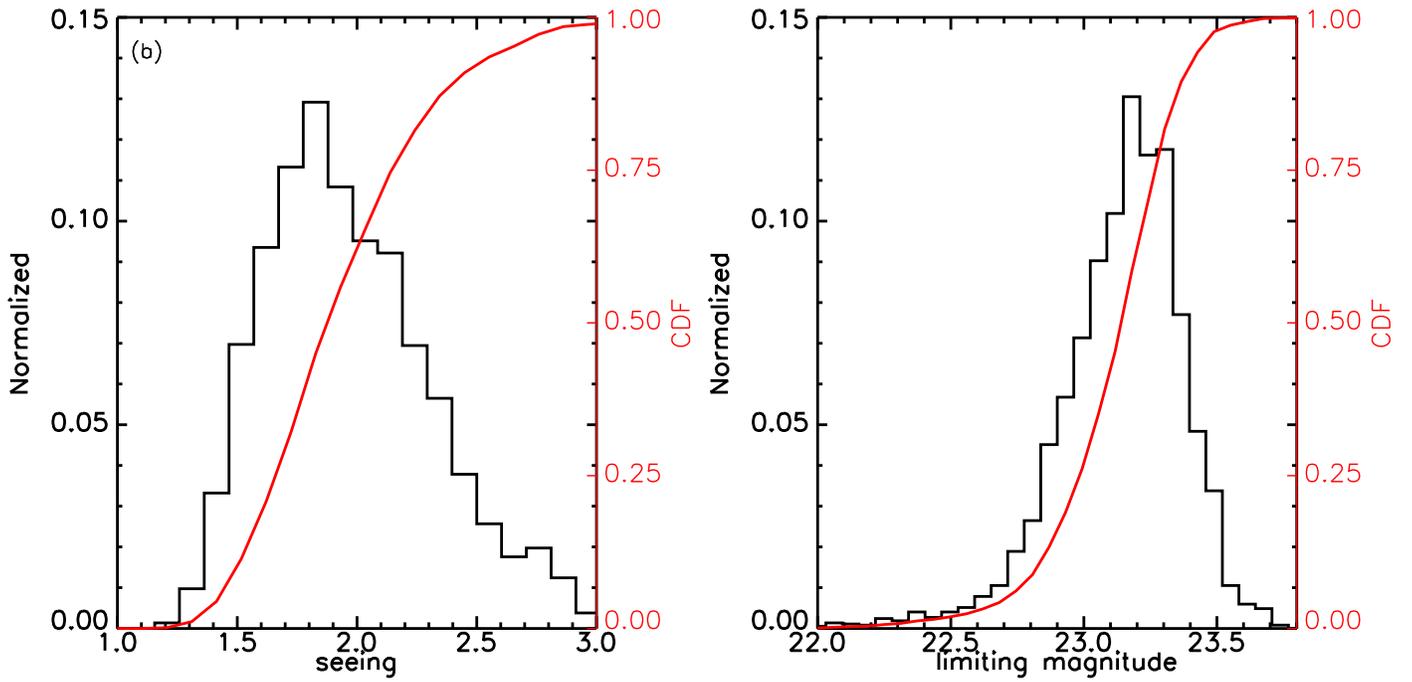}
	\caption{(a): seeing distribution in arcsec (b) magnitude limit distribution. The curves are the 
	cumulative distributions. \label{fig6}}
\end{figure*}

The overall distributions of the $u$-band Galactic extinction, sky brightness, seeing, and 
limiting magnitude are presented in Figure \ref{fig5}. The median $u$ band Galactic extinction 
is about 0.096 mag, where the extinction map comes from \citet{sch98} using the reddening law of 
\citet{car89}. Some regions with very high extinctions are not included in the SDSS footprint. Most 
observations were taken on dark nights, while a few of them were taken at gray time as seen 
in Figure \ref{fig5}b. The median seeing is about 2\arcsec.0. About 90\% of the footprint has 
seeing better than 2\arcsec.5. The $u$-band seeing is usually larger than that in redder 
bands. The typical $r$-band seeing on Kitt Peak is about 1\arcsec.7. The limiting magnitude shown 
in Figure \ref{fig5}d is estimated by using the SExtractor automatic magnitude measurements of 
5$\sigma$ point sources. The median limiting magnitude is about 23.2 mag. About 98.3\% of the footprint has 
a depth fainter than 22.5 mag. The histograms of the seeing and limiting magnitudes and 
their cumulative distributions are also shown in Figure \ref{fig6}.

\section{Catalog of Star Proper Motions} \label{sec5}
\citet{pen15} used a novel method from \citet{qiz15} to determine the absolute proper motions of 
 detected objects in SCUSS single-epoch images. Based on data from the SCUSS (2010--2013) and 
the Guide Star Catalog \uppercase\expandafter{\romannumeral2} \citep{las08} (1950--2000), the 
absolute proper motions of $\sim$ 8 million objects  were derived. A great deal of effort was 
put into correcting the position-, magnitude-, and color-dependent systematic errors. 

\begin{figure}
	\includegraphics[width=0.5\textwidth]{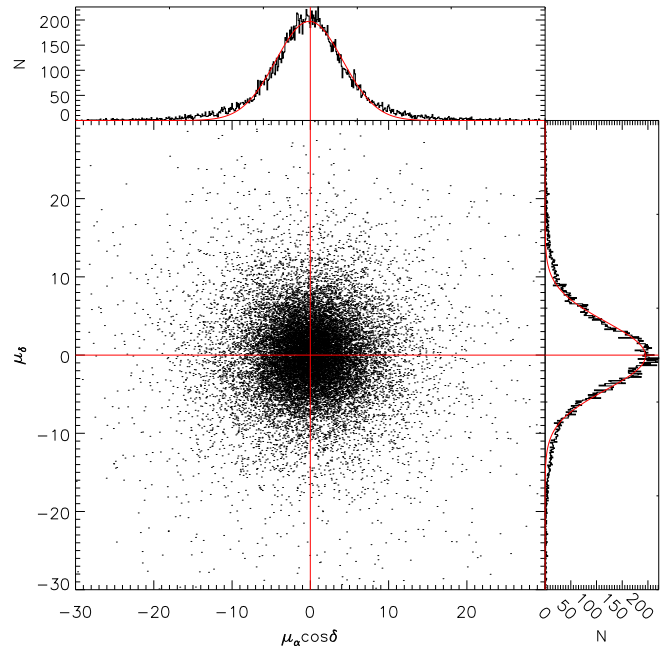} 
	\caption{The accuracy distribution of the proper motion derived by the SDSS spectroscopically confirmed quasars. The cross line show the coordinate origin. The red curves in the right and top panels are the Gaussian fits to the distributions of $\mu_{\alpha}\mathrm{cos}\delta$ and $\mu_{\delta}$, respectively. The units are $\mathrm{mas\ year}^{-1}$. \label{fig7}} 
\end{figure}

Quasars are distant and regarded to have no proper motion. The accuracy of our proper motions 
is estimated by using the spectroscopically confirmed quasars identified by the SDSS. Figure \ref{fig7} 
displays the distributions of the calculated proper motions of these quasars in  R.A.
($\mu_{\alpha}\cos\delta$) and decl. ($\mu_{\delta}$). The systematic errors (or the average) are 
about $-0.11$ and $-0.02 \mathrm{mas\ year}^{-1}$, and the corresponding random errors (or the standard deviation) 
are about 4.90 and 4.93 $\mathrm{mas\ year}^{-1}$ for $\mu_{\alpha}$ cos $\delta$ and $\mu_{\delta}$, respectively. The 
gaussian fitted random errors are 4.27 and 4.35 $\mathrm{mas\ year}^{-1}$. The random error increases with the magnitude 
from about 3 $\mathrm{mas\ year}^{-1}$ at $u = 18.0$ mag to about 7 $\mathrm{mas\ year}^{-1}$ at $u = 22.0$ mag. The SCUSS proper motions 
are compared with those in the SDSS catalog, which shows a high consistency. The typical dispersion 
of the proper motion between the SCUSS and SDSS is about 5 $\mathrm{mas\ year}^{-1}$.  Table \ref{tab4} shows the 
columns in our proper motion catalog. 

\begin{table}
\scriptsize
\centering
\caption{Columns in the catalog of star proper motions \label{tab4}}
\begin{tabular}{lll}
\hline
\hline
Field & Data type & Meaning \\
\hline
ID         & LONG        &   Object ID \\
RA         & FLOAT       &   R.A. in J2000 (degree)\\
DEC        & FLOAT       &   Decl. in J2000 (degree)\\
sigRA      & FLOAT       &   Error of RA (mas)\\
sigDEC     & FLOAT       &   Error of DEC (mas)\\
PMRA       & FLOAT       &   Proper motion in right ascension multiplied by cos($\delta$) ($\mathrm{mas\ year}^{-1}$) \\
PMDEC      & FLOAT       &   Proper motion in declination ($\mathrm{mas\ year}^{-1}$)\\
sigPMRA    & FLOAT       &   Error of proper motion in R.A.\\
sigPMDEC   & FLOAT       &   Error of proper motion in decl.\\
MAG        & FLOAT       &   SCUSS automatic magnitude\\
TYPE       & INTEGER     &   Star/galaxy classification (0 for star; others for galaxy)\\
OBSNUM     & INTEGER     &   Number of epoches\\
MeanEpoch  & FLOAT       &   Mean epoch\\
MinEpoch   & FLOAT       &   Minimum epoch\\
MaxEpoch   & FLOAT       &   Maximal epoch\\
\hline
\end{tabular}
\end{table}

\section{Conclusions} \label{sec6}
The SCUSS survey was a collaborative program between the National Astronomical Observatories 
of China and the Steward Observatory. It used the 2.3m Bok telescope and wide-field 90Prime 
camera to survey the north part of the south Galactic cap in SDSS $u$ band. The observations 
were completed in 2013 and covered about 5000 deg$^2$. This paper presents the data release 
of about 4000 deg$^2$, 92\% of which is covered by the SDSS. 

We have summarized the survey and data reduction in this paper which can be referred to \citet{zou15a} 
and X. Zhou et al. (2015, in preparation) for more details. The data products include calibrated 
single-epoch images, stacked images, photometric catalogs. The catalogs contain the photometry 
of both SCUSS detected sources and objects in SDSS catalogs and provide magnitude 
measurements on stacked images and co-added magnitudes from measurements on 
single-epoch images. The SDSS information are also included in the catalogs with a 2\arcsec matching 
error. We have analyzed the data quality, such as the sky brightness, seeing, and magnitude limit. The 
median limiting magnitude (5$\sigma$) is about 23.2 mag, which is $\sim$1.2 mag deeper than 
the SDSS $u$ band. We also release a catalog of star proper motions of about 8 million objects 
derived by \citet{pen15}.  The data and documentations can be accessed through the SCUSS data 
release website \footnote{\url{http://batc.bao.ac.cn/Uband/data.html}}. In this website, 
the images and catalogs can be retrieved either by using query forms (developed by the Chinese 
astronomical data center) or directly through the data directory trees.

\acknowledgments
This work is supported by the National Natural Science Foundation of China (NSFC, Nos. 11203031, 11433005, 11073032, 11373035, 11203034, 11303038, 11303043) and by the National Basic Research Program of China (973 Program, Nos. 2014CB845704, 2013CB834902, and 2014CB845702). Z.Y.W. was supported by the Chinese National Natural Science Foundation grant No. 11373033. 

The SCUSS is funded by the Main Direction Program of Knowledge Innovation of Chinese Academy of Sciences (No. KJCX2-EW-T06). It is also an international cooperative project between National Astronomical Observatories, Chinese Academy of Sciences, and Steward Observatory, University of Arizona, USA. Technical support and observational assistance from the Bok telescope are provided by Steward Observatory. The project is managed by the National Astronomical Observatory of China and Shanghai Astronomical Observatory. Data resources are supported by Chinese Astronomical Data Center (CAsDC).

SDSS-III is managed by the Astrophysical Research Consortium for the Participating Institutions of the SDSS-III Collaboration including the University of Arizona, the Brazilian Participation Group, Brookhaven National Laboratory, Carnegie Mellon University, University of Florida, the French Participation Group, the German Participation Group, Harvard University, the Instituto de Astrofisica de Canarias, the Michigan State/Notre Dame/JINA Participation Group, Johns Hopkins University, Lawrence Berkeley National Laboratory, Max Planck Institute for Astrophysics, Max Planck Institute for Extraterrestrial Physics, New Mexico State University, New York University, Ohio State University, Pennsylvania State University, University of Portsmouth, Princeton University, the Spanish Participation Group, University of Tokyo, University of Utah, Vanderbilt University, University of Virginia, University of Washington, and Yale University.

Based on observations obtained with MegaPrime/MegaCam, a joint project of CFHT and CEA/IRFU, at the Canada-France-Hawaii Telescope (CFHT) which is operated by the National Research Council (NRC) of Canada, the Institut National des Science de l'Univers of the Centre National de la Recherche Scientifique (CNRS) of France, and the University of Hawaii. This work is based in part on data products produced at Terapix available at the Canadian Astronomy Data Centre as part of the Canada-France-Hawaii Telescope Legacy Survey, a collaborative project of NRC and CNRS.


\begin{thebibliography}{} 
\bibitem[Bertin \& Arnouts(1996)]{ber96} Bertin, E., \& Arnouts, S.\ 1996, \aaps, 117, 393 Data Analysis Software and Systems XX, 442, 435
\bibitem[Cardelli et al.(1989)]{car89} Cardelli, J.~A., Clayton, G.~C., \& Mathis, J.~S.\ 1989, \apj, 345, 245 
\bibitem[Comparat et al.(2015)]{com15} Comparat, J., Delubac, T., Jouvel, S., et al.\ 2015, arXiv:1509.05045 
\bibitem[Cui et al.(2012)]{cui12} Cui, X.-Q., Zhao, Y.-H., Chu, Y.-Q., et al.\ 2012, Research in Astronomy and Astrophysics, 12, 1197 
\bibitem[Gu et al.(2015)]{guj15} Gu, J., Du, C., Jia, Y., et al.\ 2015, \mnras, 452, 3092
\bibitem[Jia et al.(2014)]{jia14} Jia, Y., Du, C., Wu, Z., et al.\ 2014, \mnras, 441, 503
\bibitem[Lasker et al.(2008)]{las08} Lasker, B. M., Lattanzi, M. G., McLean, B. J., et al. 2008, AJ, 136, 735
\bibitem[Nie et al.(2015)]{nie15} Nie, J.~D., Smith, M.~C., Belokurov, V., et al.\ 2015, \apj, 810, 153 
\bibitem[Peng et al.(2015)]{pen15} Peng, X., Qi, Z., Wu, Z., et al.\ 2015, \pasp, 127, 250 
\bibitem[Qi et al.(2015)]{qiz15} Qi, Z., Yu, Y., Bucciarelli, B., et al.\ 2015, \aj, 150, 137
\bibitem[Raichoor et al.(2015)]{rai15} Raichoor, A., Comparat, J., Delubac, T., et al.\ 2015, arXiv:1505.01797
\bibitem[Schlegel et al.(1998)]{sch98} Schlegel, D.~J., Finkbeiner, D.~P., \& Davis, M.\ 1998, \apj, 500, 525
\bibitem[York et al.(2000)]{yor00} York, D.~G., Adelman, J., Anderson, J.~E., Jr., et al.\ 2000, \aj, 120, 1579 
\bibitem[Zou et al.(2015a)]{zou15a} Zou, H., Jiang, Z.-J., Zhou, X., et al.\ 2015a, \aj, 150, 104 
\bibitem[Zou et al.(2015b)]{zou15b} Zou, H., Wu, X.-b., Zhou, X., et al.\ 2015b, \pasp, 127, 94


\end{thebibliography}
\end{document}